# Study on Antibody-Virus Interaction using Molecular Dynamics: Two Dimensional Simulation on Immunoglobulin Reaction against Human Papillomavirus

L. Haris[1], S. Suhandono[2], S. N. Khotimah[1], F. Haryanto[1], and S. Viridi[1, 3]

[1]Physics Department, Nuclear Physics and Biophysics Research Division, Institut Teknologi Bandung, Bandung, Indonesia

[2]Microbiology Department, Genetics and Molecular Biotechnology Research Division, Institut Teknologi Bandung, Bandung, Indonesia

[3]Computational Science, Faculty of Mathematics and Natural Sciences, Institut Teknologi Bandung, Bandung, Indonesia

*Abstract*— **Human Papillomavirus (HPV) has been known as one of the cause of virus-induced cancer such as cervical cancer and carcinoma. Among other types of cancer, this type has higher chance in being prevented earlier. The main idea is to eradicate the virus as soon as it enters the body by marking it with antibodies; signaling the immune system to dispose of it. However, the antibodies must be trained to recognize the virus. They can be trained by inserting an object similar to the virus allowing them to learn to recognize and surround the inserted object. In response to this, molecular dynamics simulation was chosen to study the antibody-virus interaction. In this work, two-dimensional case that involves HPV and immunoglobulin (Ig) was studied and observed. Two types of objects will be defined; one stands for HPV while another stands for antibodies. The interaction between the two objects will be governed by two forces; Coulomb force and repulsive contact force. Through the definition of some rules and condition, the antibodies' motion was observed. The influence of antibody concentration, and the antibody's type and their appearance sequence were observed both quantitatively and visually.**

## I. Introduction

Antibody has been known to react to any intruding organisms such as bacteria or viruses. These antibodies, also known as immunoglobulin (Ig), have five isotypes; IgG, IgM, IgA, IgD, and IgE [1]. The classification is based on the number of monomers that constitute them. This mechanism has been utilized as a preventive effort in fighting off many virus-induced diseases including cancer. Aside from carcinogen-induced cancer, there are also some that are induced by viruses such as cervical cancer and carcinomas. In the early stage, these viruses are dealt like any other intruding organisms; marked by Ig when they intrude our body. Depending on the intruding site, the antibody isotypes may vary. Once marked, the immune system will recognize the intruders, and ultimately disposed the body of them. Hence, the virus-induced cancer has an advantage in being able to be prevented earlier; even before the invasion starts. However, despite the convenience, the antibody must be trained to recognize the virus. The training can be done by inserting an object similar to the virus allowing them to recognize and cover it. Hence, in accordance with the previous statement, a theoretical study and observation on antibody-virus interaction was carried out. This work aims to study the case of Human Papillomavirus (HPV) intrusion into the body that will be covered by three isotypes; IgG, IgM, and IgA. The results are expected to give some insight on the antibodies distribution on the virus' surface.

## II. Methods

Both antibody and virus consist of proteins which enable us to model them as charged object [2, 3]. In terms of dimensions, the actual shape of HPV is spherical; leaving antibodies that need to be simplified. The basic form of antibody is monomer which has Y-like shape, consisting heavy chain and light chain. IgG, IgD, and IgE are monomers while IgM and IgA are pentamer and dimer respectively [1]. However, among these isotypes, only IgG, IgM, and IgA are involved in the virus' marking; particularly IgA which is involved since we are working with HPV [4, 5]. The simplification of the Ig is based on the constant region as illustrated in Figure 1. The Ig will be given spherical shape whose diameter represents that of the constant region. Finally, the value of the necessary properties of both antibodies and virus are provided within Table 1.

Table 1 The properties of IgG, IgM, IgA, and HPV obtained from numerous literatures.

|  | Mass (kDa)* | Diameter (nm) | Charge (nC) |
|---|---|---|---|
| HPV | 880 | 55 | $1.6\times10^{-10}$ |
| IgG | 160 | 10 | $-1.6\times10^{-10}$ |
| IgM | 320 | 25 | $-1.6\times10^{-10}$ |
| IgA | 939 | 20 | $-1.6\times10^{-10}$ |

*This value was not used in the simulation. These numbers were converted using the size ratio between HPV and the respective antibody isotypes.

Several values are provided by some pharmaceutical companies while others are the results of certain measurement techniques [6]. Due to the numerical limitation, the value of the objects' mass are converted based on the size ratio. These values along with the other simulation parameters are given in Table 2. On the other hand, all of the objects were given charge whose magnitude is the same as that of electron.

Table 2 The value of the objects' mass used in the simulation along with the other simulation parameters.

|  | IgG | IgM | IgA |
|---|---|---|---|
| Mass (ng) | 0.1 | 0.1 | 0.1 |
| HPV Mass (ng)* | 0.55 | 0.275 | 0.939 |
| *Relative to the antibody's | | | |
| $k_q$ | $9 \times 10^{20}$ | | |
| $k_r$ | $2 \times 10^{-5}$ | | |
| $k_v$ | 20 | | |
| $\Delta t$ | $10^{-5}$ | | |

In this work, the problem was simplified into two dimensions, reducing sphere into circle. The scheme used in the simulation involves one virus and several antibodies. These objects are governed by forces similar to the one used in the previous work with the absent of Stokes force [7]. It is due to the assumptions that the flow of any fluids surrounding these objects doesn't significantly affect the antibodies' movement.

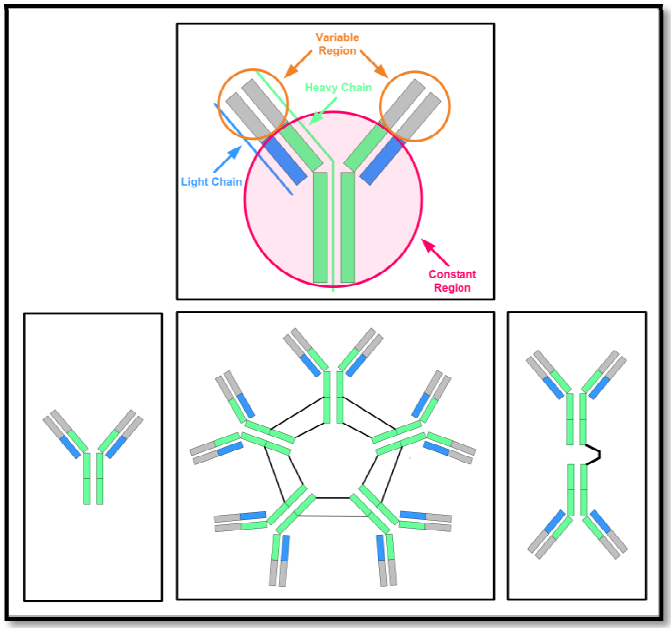

Figure 1 The illustration on immunoglobulin (Ig) and three of its isotypes; IgG, IgM, and IgA respectively (below, from left to right). The constant region of Ig is used to determine the Ig diameter. (Pictures are drawn based on the illustration provided within Kyowa Hakko Kirin Co.,Ltd website). Further details on Ig can also be obtained in the following literature [1].

Hence, leaving two forces that governs their motion which is attractive force (Coulomb force) and repulsive contact force. Both the explanation and formulation of these forces are already provided in the previous work [7, 8]. The HPV was set to a fixed position while the antibodies' were given arbitrary position. Afterward, molecular dynamics was used to obtain the number of antibodies that were successfully bound to the HPV. The antibodies that were attached to the HPV were called bounded antibodies. Finally, it is also worth noted that all of the quantities are not in SI unit. In place of SI unit, all quantities were determined in nano scale.

## III. RESULT AND DISCUSSION

All simulation parameters were chosen subjectively using a scheme that involves one HPV and one antibody. They were chosen such that it will not require huge amount of computation time. Coulomb constant was determined with this idea in mind since there is no other force that exists within the system. It was chosen to give a considerable gradual speed up to the antibody. Furthermore, through the same scheme, it was found that the constant should have an order of more than $10^{10}$. However, due to the magnitude of the objects' charge, the corresponding value within Table 2 was used. On the other hand, damping constant, which was a part of attractive contact force, was chosen to quickly reduce the Coulomb force to zero once the antibody touched another object; hence stopping the movement. However, this approach comes with a price since the antibody that come in contact with bound antibody will still be attracted to HPV thus potentially producing overlap between antibodies.

Due to the penalty, the determination of whether or not the antibody was deemed as being bounded to the HPV was done in two steps. The first step was done by analyzing the raw data. Any antibodies that were able to reach the minimum distance between an antibody and HPV will be considered as being bounded. The next step is to visually check whether or not these bounded antibodies overlap with each other. If the overlap is small enough, then it can be considered as geometrical error due to the simplification; hence the corresponding antibodies would still be considered bounded. However, if the overlap is dominant, then the later antibody will be left out while the former one will be considered as being bounded. This selection step was used to rule out which antibody was bounded and which was not in the following results.

Table 3 List of treatments done in the simulation.

| | Variations | Value | | |
|---|---|---|---|---|
| Five repetitions | Antibody concentration $\left( N_{Ig} / N_{HPV} \right)$ | 5…10 | | |
| | Antibody types* | IgG | IgM | IgA |

*The properties were already given within Table 1 and Table 2.

The initial position of the antibodies was arbitrarily chosen, and the calculation for the next position was done one by one.

In other words, once an antibody was simulated, the rest will be considered as non-exist except those who were already considered as being bounded. This scheme will prevent antibodies to repel each other at the start of the simulation.

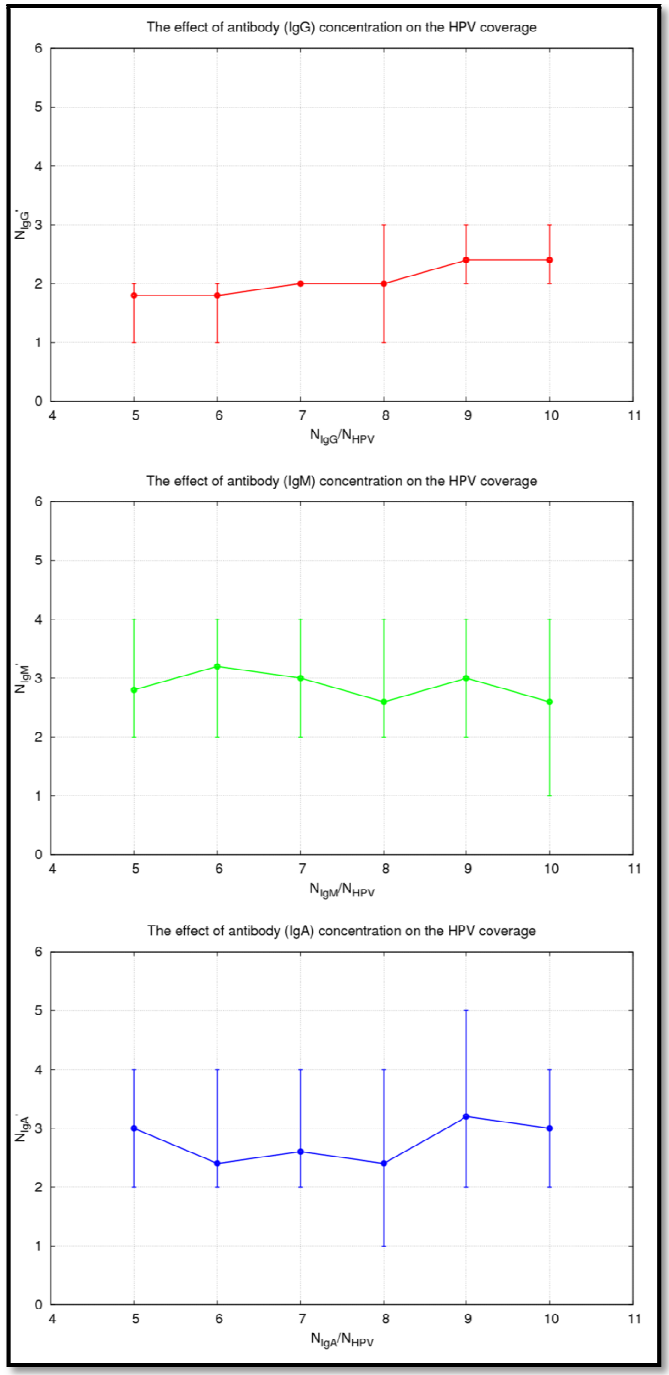

Figure 2 The number of antibodies that are considered as being bounded to HPV $(N_{Ig}')$ over five repetitions. Data deviation for (upper to lower) IgG, IgM, and IgA is also provided.

Moreover, since the initial position was chosen arbitrarily, repetition was necessary. In this work, five repetitions were done for each type of antibody and each variation on the ratio between the number of antibodies and the number of virus. These treatments were listed within Table 3 while the simulation results can be seen in Figure 2.

The error graphs show us that initial position holds significant importance of producing bounded antibodies. Among these, only IgG that show the increment of the bounded antibodies due to the increase of the ratio between the number of antibodies and the number of HPV. The reason behind the other two graphs lies with the second part of the determination step. This argument can be easily understood by observing Figure 3.

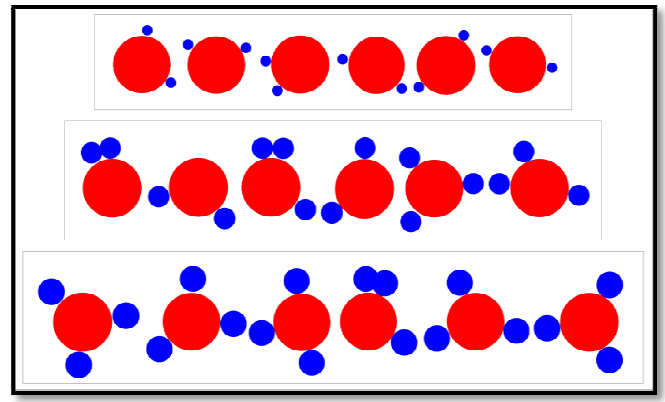

Figure 3 Visualization sample from the simulation data. The blue circles and red circles represent bounded antibodies and HPV respectively. The visualizations are divided into three blocks. The first one (uppermost block) is the visualization for IgG, the middle one for IgA, and the last one for IgM. For each block, the visualization for all concentration ratios (left to right) was given.

All of the visualizations do not reveal dominant overlap since they were eliminated in the selection step. However, some exhibit slight overlap which is still within toleration, and is considered as geometrical error.

There is also another factor beside initial position that affect the number of antibody that can bind itself to the HPV; the appearance sequence. Typically, the abundance of IgA is more than IgM while being lesser than IgG. In terms of the order of appearance, IgM is the first one to be produced while the immune system prepares the IgG. On the other hand, IgA will be dispatched near the infection site which, in this case, is the cervix [1].

Table 4 The list of appearance sequences and the number of antibodies involved in the simulation.

| Sequence1 | IgM | IgG | IgA |
|---|---|---|---|
| Number of antibodies | 2 | 4 | 3 |
| Sequence 2 | IgM | IgA | IgG |
| Number of antibodies | 2 | 3 | 4 |

The sequence and the number of antibodies is provided within Table 4. IgM will always be dispatched first; leaving IgG and IgA whose order of appearance can be switched. The proof of the effect of appearance sequence is given within Figure 4.

The initial position of the antibodies was determined randomly, but the position used in the two sequences was the exactly the same. Note that not only the position of both IgG and IgA differ for both sequences, but the possibility of them for not being able to be bounded to HPV also differ.

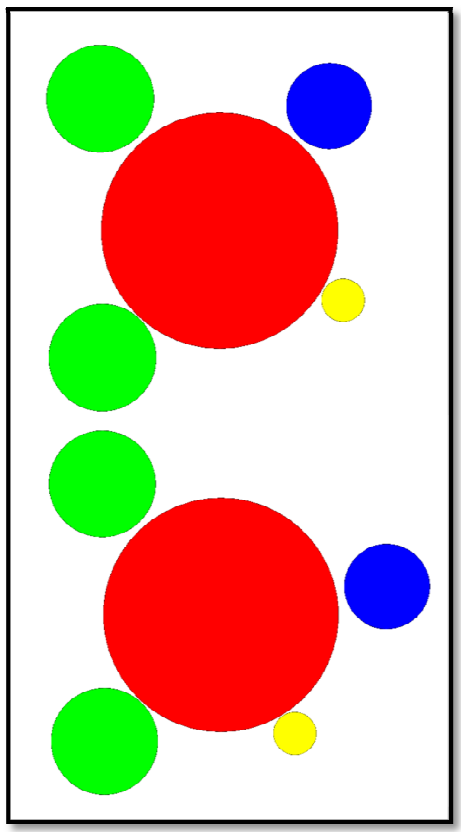

Figure 4 Visualizations on the effect of appearance sequence on antibody distribution. Circles with the color of red, green, blue, and yellow are HPV, IgM, IgA, and IgG respectively. The upper part is the visualization for sequence 1 while the lower part is the visualization for sequence 2.

It was also found that the magnitude of the HPV should differ from that of antibody to avoid antibodies from being thrown away. It may happen due to the repulsive force from the surrounding bounded antibodies. It was highly likely that the results in Figure 2 and Figure 3 were also influenced by this condition. Aside from the previously mentioned argument concerning initial position, it also explains the existence of some spacious surface that has not yet occupied as seen in Figure 3.

## IV. Conclusion

The proposed model was able to give out the desired output for IgG although the penalty due to the time-cutting approach gave out different results for the other types; IgM and IgA. It was found out that both initial position and appearance sequence of the antibodies are significantly important in determining the antibody distribution on the HPV surface. However, quantitative proof cannot yet be provided since it was not done in the current work.

## Acknowledgement

This work is partially supported by Riset Inovasi KK (RIK) ITB in year 2013 with contract number 241/I.1.C01/PL/2013 and 248/I.1.C01/PL/2013 enabling the corresponding presentation on the 11[th] South East Asian Congress of Medical Physics and the 13[th] Asia Oceania Congress of Medical Physics held in Singapore.

Address of corresponding author:
    Author: Luman Haris
    Institute: Institut Teknologi Bandung
    Street: Jl. Ganesha no 10
    City: Bandung
    Country: Indonesia
    Email: ign_lumanharis@hotmail.com